\documentstyle[12pt]{article}
\newcommand{\Dbar}{\not{\!{\!D}}}

\newcommand{\mapright}[1]{\smash{\mathop{\hbox to 1cm{\rightarrowfill}}\limits_{#1}}}

\def\gappeq{\mathrel{\rlap {\raise.5ex\hbox{$>$}}
{\lower.5ex\hbox{$\sim$}}}}

\def\lappeq{\mathrel{\rlap{\raise.5ex\hbox{$<$}}
{\lower.5ex\hbox{$\sim$}}}}

\begin{document}
\input epsf \renewcommand{\topfraction}{0.8}
\pagestyle{empty}
\begin{flushright}
{SU-ITP-01/47}
\end{flushright}
\vspace*{5mm}
\begin{center}
\Large{\bf Constraining the primordial spectrum of metric perturbations 
 from gravitino and moduli production} \\
\vspace*{1cm} 
\large{\bf Antonio L. Maroto}$^{*}$  \\
\vspace{0.3cm}
\normalsize 
Physics Department, Stanford University, \\
Stanford CA 94305-4060, USA

\vspace*{1cm}  
{\bf ABSTRACT} \\ \end{center}
\vspace*{5mm}
\noindent
We consider the production of gravitinos and moduli fields from quantum
vacuum fluctuations induced by the presence of scalar metric
perturbations at the end of inflation. We obtain the corresponding occupation
numbers, up to first order in perturbation theory, in terms of the power
spectrum of the metric perturbations. We compute the limits imposed by 
nucleosynthesis on the spectral index $n_s$ for different models
with constant $n_s$. The results show
that, in certain cases,
such limits can be as strong as $n_s<1.12$, which is more stringent than 
those coming from primordial black hole production.

\vspace*{0.5cm}


\vspace{4cm}

\noindent 
\rule[.1in]{8cm}{.002in}

\noindent
$^{*}$ On leave of absence from Dept. F\'{\i}sica 
Te\'orica, Universidad Complutense de Madrid,\\
 28040, Madrid, Spain

\noindent E-mail: maroto@itp.stanford.edu
\vfill\eject

\setcounter{page}{1}
\pagestyle{plain}

\newpage

\section{Introduction}

In the framework of the inflationary cosmology, density perturbations are
generated when metric and inflaton quantum fluctuations 
become super-Hubble sized during
inflation and reenter the horizon as classical fluctuations in the radiation
or matter dominated eras \cite{brand}. 
Typically, the scale corresponding to the present 
horizon size ($H_0^{-1} \sim 3000$ Mpc) 
left the horizon about 60 e-folds before the 
end of inflation,
and therefore one expects a primordial spectrum of 
metric perturbation spanning a range of 
scales from thousands of megaparsecs down to $e^{-60}/H_0 \sim 10^{-23}$ 
Mpc. The current observations of cosmic microwave background anisotropies and
large scale structure have started providing us with precise measurements
of the primordial spectrum at large scales (1000 - 1 Mpc). Smaller scales 
are much more difficult to probe since their evolution  has already 
become non-linear or the corresponding fluctuations  
 have been erased by various damping 
 effects in the early universe. However, still there is a window open
to those scales which is provided by primordial black holes.
When a perturbation crosses inside the
horizon with relative large amplitude $1/3\leq \delta \leq 1$, the
perturbed region stops expanding and starts collapsing 
eventually forming a black hole \cite{Carr,Liddle,Green}. 
If such black holes do not evaporate fast enough they could overclose
the universe and, therefore, limits on their primordial abundance 
 are required. If they do evaporate by the present time, 
limits are still needed in order for the evaporation 
products  do not 
disrupt primordial nucleosynthesis. Such limits  can be translated 
into bounds on the power spectrum, which are particularly useful
to constrain models of inflation in which perturbations
grow at smaller scales, i.e.,  when the corresponding spectral index is 
larger than that of  Harrison and Zeldovich  $n_s=1$.
  
Apart from the generation of primordial black holes, metric perturbations
at small scales can have other interesting effects, which we will explore in
this paper.  The presence of metric inhomogeneities is known to  break
 the conformal flatness of the Friedmann-Robertson-Walker (FRW) metric. 
In fact, such effect is more important as we move to smaller and smaller 
scales, and thus, in principle, one expects to find a strong deviation from
conformal flatness right at the end of inflation, provided the smallest
scales reenter the horizon with sufficiently large amplitude.
However, we will find that even perturbations with very
small amplitude can have important cosmological effects.
One of the physical consequences of conformal 
non-invariance is the possibility of
creating particles from vacuum fluctuations
\cite{Zeldovich,Verdaguer}. Recently 
\cite{magnetic,basset1} this
mechanism has been applied to the generation of large scale magnetic fields
after inflation. Also in \cite{Bassett}, the 
production of fermions from metric perturbations
during preheating has been considered. Other sources for fermion
production after inflation have been studied in \cite{mama1}.
In this paper 
we will be interested in the creation of
a different kind of relics which can also have  important consequences in
cosmology, namely gravitinos and moduli fields. 

In supergravity theories, the gravitino is the spin-$3/2$ superpartner
of the graviton field. Its couplings to the rest of matter fields are
typically  
supressed by the Planck mass scale, which implies that gravitinos can live
very long and even decay after nucleosynthesis destroying the nuclei
created in this period. A similar effect is due to the presence of moduli
fields, corresponding to the radii of extra dimensions in higher dimensional
theories. This imposes very strong contraints on the primordial abundance
of this kind of particles, thus typically we have 
$n/s<10^{-12}-10^{-14}$ for masses of the relics in the range 
$m=10^{3}-10^{2}$ GeV \cite{Ellis,Ellis2,LiEll,Sarkarrep,Moroi}. 
Traditionally, these constraints are translated into
upper limits on the  reheating temperature of the universe. In 
order to avoid
their thermal overproduction we need  $T_R<10^8-10^9$ GeV. 
More recently, 
non-thermal
production of gravitinos and moduli during preheating has been considered
\cite{mama,LiKa,Riotto1,mape,Linde3,LiKa2}.
In those works, it has been shown how the coherent oscillations of the
inflaton field after inflation could produce a large amount of relics
that could conflict in some cases with the above limits. In fact,
in the case of moduli fields, the limits  imposed on the reheating
temperature can be as strong as  $T_R<100$ GeV, 
if the  gravitational or the production during preheating 
are taken into account \cite{Linde3}.
In this work, we will show how in certain models of inflation, 
metric perturbations can also give rise to a relevant amount of gravitinos
and moduli. In those cases, the nucleosynthesis limits discussed above
impose stringent constraints on the perturbations power spectrum.   

The plan of the paper goes as follows. In Section 2, we consider the gravitino
equations of motion in an inhomogeneous background and obtain their
 perturbative solutions. In Section 3, using the Bogolyubov technique,
 we compute the total number of gravitinos
produced as a function of the power spectrum  and compare the 
results with the nucleosynthesis bounds. Section 4 is devoted to a 
similar analysis but using moduli
fields. Section 5 contains the main conclusions of the work and finally we 
have summarized some useful formulae in the Appendix.

\section{Gravitino field equations in inhomogeneous backgrounds}

The massless Rarita-Schwinger equation is conformally invariant, which implies 
that gravitinos are not produced from vacuum fluctuations in a FRW background.
In the case of massive gravitinos in FRW, they can be produced
either  gravitationally
\cite{Lemoine} 
or due to the time dependence of the mass during preheating 
\cite{mama,LiKa,Riotto1}. However,
since in this work we are interested in the effect of metric perturbations,
we will consider that the mass of the gravitino is constant and very small
and effectively we can neglect it in the calculations. This is 
a good approximation for models in which the gravitino
mass is smaller than the Hubble parameter after inflation, 
since gravitinos will be produced with energies much higher than their mass. 
 For simplicity, we will also consider only the production of
helicity $\pm 3/2$ particles which is technically simpler.
The helicity $\pm 1/2$ production exhibits additional subtelties related to 
the fact that their equations of motion are only consistent if the 
gravitational background
is a solution of the  supergravity equations of motion. In order to get
consistent inflationary supergravity models, one usually requires the
introduction of several chiral superfields. In that case, the equations
of motion for the helicity $\pm 1/2$ gravitino cannot be decoupled from
those of the fermionic components of the superfields
\cite{LiKa2}. In addition, in the context of the production during
preheating, 
if the inflationary sector is decoupled from the sector responsible 
for the present supersymmetry breaking, it is the fermionic 
partner of inflaton what is produced more abundantly rather than
the longitudinal gravitino \cite{Peloso}.  These points make the 
calculations more involved. However, in our case, we do not expect any 
new physical 
effects coming
from helicity $\pm 1/2$, since     
on physical
grounds, the production mechanism is purely gravitational and 
the helicity $\pm 1/2$ components will be produced with similar abundance
to the helicity $\pm 3/2$ states. 

Let us then consider the massless Rarita-Schwinger equation in an external 
gravitational background:
\begin{eqnarray}
\epsilon^{\mu\nu\rho\sigma}\gamma_5\gamma_\nu D_\rho\psi_\sigma=0.
\label{RS}
\end{eqnarray}
where $\psi_\sigma$ is a  Majorana spinor
satisfying $\psi_\sigma=C\bar\psi_\sigma^T$ 
with $C=i\hat\gamma^2\hat\gamma^0$ the
charge conjugation matrix. The covariant derivative is given by:
\begin{eqnarray}
D_\mu\psi_\sigma=(\partial_\mu+\Gamma_\mu)\psi_\sigma-
\Gamma^\lambda_{\mu\sigma}\psi_\lambda
\end{eqnarray}
and the spin connection  by: $\Gamma_\mu=-\frac{1}{8}
\Gamma_\mu^{ab}[\hat\gamma_a,\hat\gamma_b]$. 
With these definitions, we have 
$D_\mu\gamma_\nu=0$ and because of the totally antisymmetric tensor, 
the Christoffel symbols
do not contribute to (\ref{RS}). As usual, latin indices $a,b,\dots$ refer to
the tangent-space tensors, whereas greek ones $\mu,\nu,\dots$ are used
for curved background objects. The different gamma matrices are related
by $\gamma_\mu=e^a_\mu\hat\gamma_a$, where the vierbein satisfies:
$e^a_\mu e^b_\nu g^{\mu\nu}=\eta^{ab}$, with $\eta_{ab}$ the
Minkowski space metric.

For the background metric we will take the following form:
\begin{eqnarray}
g_{\mu\nu}=g_{\mu\nu}^{0}+h_{\mu\nu}
\label{metric}
\end{eqnarray}
where
\begin{eqnarray}
g^{0}_{\mu\nu}dx^\mu dx^\nu=a^2 (\eta)(d\eta^2-\delta_{ij}dx^idx^j)
\end{eqnarray}
is the flat FRW metric in conformal time and
\begin{eqnarray}
h_{\mu\nu}dx^\mu dx^\nu=2\Phi a^2(\eta)(d\eta^2+\delta_{ij}dx^i dx^j)
\end{eqnarray}
is the most general form of the linearized  scalar metric perturbation in the
longitudinal gauge and where it has been assumed that the spatial
part of the energy-momentum tensor is diagonal, as indeed happens in
the inflationary or perfect fluid cosmologies \cite{brand}. In this expression
$\Phi(\eta,\vec x)$ is the gauge invariant gravitational potential.

Contracting equation (\ref{RS}) with $\gamma_\lambda\gamma_\mu$ we get:
\begin{eqnarray}
2iD_\lambda(\gamma^\sigma\psi_\sigma)-2i\Dbar\psi_\lambda=0
\label{eom}
\end{eqnarray}
As commented before we are asumming that metric perturbations are classical
perturbation which are produced during inflation. In this sense, it is
a good approximation to consider that $\Phi$ vanishes asymptotically in
the past. Since we are interested in the smallest scale
perturbations, which are going to produce the leading effects, we
will consider only those perturbations that reenter the 
horizon during the radiation era. In this case, once they reenter
they start oscillating with damped amplitude, for that reason, we will
also take $\Phi\rightarrow 0$ when $\eta\rightarrow \infty$. 
The vanishing of the perturbations  allows us to define conformal 
vacuum states in the asymptotic regions.
Thus, in those regions, the above equation reduces to
a Dirac-like equation $i\Dbar^{(0)} \psi_\mu^{(0)}=0$
(see \cite{mama}) where the index $(0)$ denotes the
unperturbed object. The corresponding positive frequency 
solution with momentum $\vec p$ and helicity $\lambda=\pm 3/2$ can be 
written as:

\begin{eqnarray}
\psi_\mu^{(0)p,\pm}(x)&=&a^{-1/2}(\eta)\tilde\psi_\mu^{(0)\pm}(\vec p,\eta)
e^{i\vec p\vec x}\nonumber \\
&=&
\frac{1}{\sqrt {2pVa(\eta)}}u(\vec p,\pm)
\epsilon_\mu(\vec p,\pm)
e^{i\vec p\vec x-ip\eta}
\label{zero}
\end{eqnarray}
where the polarization vectors are given by
\begin{eqnarray}
\epsilon_\mu(\vec p, +)&=& \frac{1}{\sqrt{2}}
(0,\cos \theta \cos\phi-i\sin \phi,
\cos\theta \sin\phi+i\cos\phi,-\sin\theta)\\
\epsilon_\mu(\vec p,-)&=&-\frac{1}{\sqrt{2}}
(0,\cos \theta \cos\phi+i\sin \phi,
\cos\theta \sin\phi-i\cos\phi,-\sin\theta)
\end{eqnarray}
with $p^{\mu}=(p, p\sin \theta \cos \phi,
p\sin \theta \sin \phi, p\cos \theta)$ and the normalization  
 $\epsilon_\mu^*(\vec p,m)\epsilon^\mu(\vec p,n)=-\delta_{mn}$,
$p^\mu\epsilon_\mu(\vec p,m)=p^\mu\epsilon_\mu^*(\vec p,m)=0$.
The helicity $r,s=\pm 1/2$ spinors are chosen such that they satisfy: 
$u^\dagger(\vec p,r)u(\vec p,s)=2p\, \delta_{rs}$. Here we are working
in a finite box with comoving volume $V$ and we will take the infinite volume
limit at the end of the calculations.
Notice that the above solutions satisfy the
additional constraints \cite{mama}:
\begin{eqnarray}
p^\mu\psi_\mu^{(0)p,\pm}(x)=0,\;\;\gamma^\mu\psi_\mu^{(0)p,\pm}(x)=0.
\label{cons}
\end{eqnarray}

We will look for perturbative solutions of  equations (\ref{eom})
 in the form:
\begin{eqnarray}
\psi_\mu=\psi_\mu^{(0)p,\lambda}+\psi_\mu^{(1)}+\dots
\end{eqnarray}
where, as mentioned above, $\psi_\mu^{(0)p,\lambda}$ is the solution of the
unperturbed equation given in (\ref{zero}).   
Expanding the  equations (\ref{eom}) up to first order in
the perturbation we find:
\begin{eqnarray}
2iD^{(0)}_\mu(\gamma^{(0)\sigma}\psi^{(1)}_\sigma)-2i\Dbar^{(0)}
\psi^{(1)}_\mu
-2i\Dbar^{(1)}\psi^{(0)}_\mu=0
\label{pert}
\end{eqnarray}
where we have used the constraint equations (\ref{cons}).
In Appendix A we have given the expressions for the perturbative
expansions of the different terms in the above equation. 
Since we are only interested in the evolution of those states which
asymptotically (in the past and in the future)
give rise to helicity $\pm 3/2$ gravitinos, we will
project these equations along those helicity  states. The projectors in the
asymptotic regions are given in Fourier space by
\begin{eqnarray}
P^\mu_{\pm 3/2}(\vec k)=\epsilon^{\mu *}(\vec k,\pm) P_{\pm 1/2}(\vec k)
\end{eqnarray}
where $P_r$ are helicity $r=\pm 1/2$ projectors satisfying 
 $P_r(\vec k)\; u(\vec k,s)=u(\vec k,s)\delta_{rs}$.
Let us assume for simplicity that the three momentum is pointing
along the z-direction $k^{\mu}=(\vert k\vert,0,0,k)$.
In this case, the projectors take a very simple form:
\begin{eqnarray}
\epsilon^{\mu *}(\vec k,\pm)=\frac{1}{\sqrt{2}}(0,1,\mp i,0)\nonumber \\
P_{\pm 1/2}(\vec k)=\frac{1}{2}(1 \pm \gamma^5\hat\gamma^0\hat\gamma^3)
\end{eqnarray}
 Notice that the
$\mu=0$ equation in (\ref{pert}) does not contribute to the 
projected equations. Using the fact that $P^\mu_{\pm 3/2}(\vec k)k_\mu \psi=0$
and $P^\mu_{\pm 3/2}(\vec k)\gamma_\mu \psi=0$ for an arbitrary spinor $\psi$,
we see that the first term in  (\ref{pert}) vanishes when projected
and we are left with:

\begin{eqnarray}
P^i_{\pm 3/2}(\vec k)\int\frac{d^3 x}{(2\pi)^{3/2}}
e^{i\vec k \vec x}(\Dbar^{(0)} \psi_i^{(1)}+\Dbar^{(1)} \psi_i^{(0)})=0
\end{eqnarray} 

Using the formulae in Appendix A we get the  
following expression for the equations of motion:

\begin{eqnarray}
0&=&\left(\hat \gamma^0\partial_0-i\hat\gamma^j k_j\right)\tilde 
\psi_{\pm 3/2}^{(1)}(\vec k,\eta)
-P^i_{\pm 3/2}(\vec k)\left[\Phi\left(
\hat \gamma^0\partial_0-i\hat\gamma^j p_j\right )\tilde\psi_i^{(0)\lambda}
(\vec p,\eta)
\right]\nonumber \\
&-&\frac{i}{2} \Phi P^i_{\pm 3/2}(\vec k)\left[(k_j+p_j)\hat\gamma^j
\tilde\psi_i^{(0)\lambda}(\vec p,\eta)\right]
\label{ecuacion}
\end{eqnarray}
where 
\begin{eqnarray}
\tilde 
\psi_{\pm 3/2}^{(1)}(\vec k,\eta)=a^{1/2}P^i_{\pm 3/2}(\vec k)
\psi_i^{(1)}(\vec k,\eta)
\end{eqnarray}
with $\psi_i^{(1)}(\vec k,\eta)$ the corresponding Fourier mode, defined
as usual by $f(\vec k,\eta)=(2\pi)^{-3/2}\int d^3 x 
e^{i\vec k \vec x}f(x)$. We have 
already defined  $\tilde\psi_i^{(0)\lambda}(\vec p,\eta)$ in (\ref{zero}).
Finally we have denoted $\Phi(\vec k+\vec p,\eta)$ 
simply by $\Phi$.

Notice that since $\tilde\psi_i^{(0)\lambda}
(\vec p,\eta)$ is a solution of the unperturbed equations of motion, the
second term in the previous equation vanishes. Finally, in order to reduce
(\ref{ecuacion}) to a standard (inhomogeneous)  
harmonic oscillator equation, we multiply by 
 $\left(\hat \gamma^0\partial_0
-i\hat\gamma^j k_j\right)$, and  we get:
\begin{eqnarray}
\left(\partial_0^2+k^2\right)\tilde\psi^{(1)}_{\pm 3/2}(\vec k,\eta)
+J_{\pm 3/2}=0
\label{perteq}
\end{eqnarray}
The spinor current is given by:
\begin{eqnarray}
J_{\pm 3/2}=\frac{1}{2}\left[(p^2+k^2-2pk_j \hat\gamma^0\hat\gamma^j)
\Phi- i\hat \gamma^0(k_j\hat \gamma^j-p\hat \gamma^0)\Phi'
\right]
P^i_{\pm 3/2}(\vec k)\tilde\psi^{(0)\lambda}_i(\vec p,\eta)\nonumber
\end{eqnarray}
where prime denotes derivative with respect to conformal time $\eta$.
Because of the presence of the inhomogeneous current, 
the initial positive frequency solution, with momentum
$\vec p$ and helicity $\lambda$ in (\ref{zero}) 
will evolve into  a linear superposition
of  positive and negative frequency modes, with different helicities and
different momenta. Thus in the asymptotic future we find:
\begin{eqnarray}
\psi_\mu^{p,\lambda}(x)&\mapright{\eta\rightarrow \infty}&
 \sum_{\lambda'=\pm}\sum_k
\left(\alpha_{pk\lambda\lambda'}\frac{u(\vec k,\lambda')
\epsilon_\mu(\vec k,\lambda')}{\sqrt{2kVa(\eta)}}
e^{i(\vec k\vec x -k\eta)}\right.
\nonumber \\
&+&\left. \beta _{pk\lambda\lambda'}\frac{u^C(\vec k,\lambda')
\epsilon_\mu^*(\vec k,\lambda')}{\sqrt{2kVa(\eta)}}
e^{-i(\vec k \vec x-k\eta)}\right)
\end{eqnarray}

In order to obtain the Bogolyubov coefficients, we need to solve 
(\ref{perteq}). Up to first order in perturbations, we have:
\begin{eqnarray}
\tilde\psi_{\pm 3/2}^{p,\lambda}(\vec k,\eta)=
\tilde\psi^{(0)p,\lambda}_{\pm 3/2}(\vec k,\eta)
+\frac{1}{k}\int_{\eta_0}^\eta J_{\pm 3/2}
\sin(k(\eta-\eta'))d\eta'
\end{eqnarray}
where $\eta_0$ denotes the initial time in the remote past 
when the perturbations were switched off.   
Comparing this expression with the above expansion, it is very simple
to get an explicit expression for the Bogolyubov coefficients \cite{Birrel}:
\begin{eqnarray}
\beta_{pk\lambda\mp}=-\frac{i}{2k\sqrt{2kV}}\int_{\eta_0}^{\eta_1}
u^{C\dagger}(\vec k,\mp)J_{\pm 3/2}\,e^{-ik\eta}
\end{eqnarray}
where $\eta_1$ denotes the present time or any other instant in time
after inflation in which the perturbations vanish again.  
Notice that the helicity projectors $P^\mu_{\pm 3/2}$
project positive frequency modes on $\pm 3/2$ helicity states, whereas
negative frequency modes are projected on $\mp 3/2$ states, that is the reason
why the $\beta$ coefficients subindex are changed with respect to those
of the current $J_{\pm 3/2}$. The total number of gravitinos with 
comoving momentum $k$ is given by:
\begin{eqnarray}
N_k=\sum_{\lambda\lambda'}
\sum_p \vert \beta_{pk\lambda\lambda'}\vert^2
\label{total}
\end{eqnarray}

\section{Gravitino production from metric perturbations}
As commented before, we will concentrate only in the effect of 
super-Hubble scalar perturbations whose
evolution is relatively simple. For single-field inflationary scenarios
we have \cite{brand}:
\begin{eqnarray}
\Phi(\vec k,\eta)=C_k\frac{1}{a}\frac{d}{d\eta}\left(\frac{1}{a}\int a^2
  d\eta\right)+ D_k\frac{a'}{a^3},
\label{perturb}
\end{eqnarray}
the second term decreases during inflation and can soon  be neglected. 
Thus, it will be useful to rewrite the perturbation 
as: $\Phi(\vec k,\eta)=C_k {\cal F}(\eta)$. 
During inflation or preheating, super-Hubble 
perturbations evolve in time, whereas they are
practically constant during radiation or matter eras. Notice that
this expression is not appropriate in two-field models of inflation, since
in that case super-Hubble gravitational fluctuations can be exponentially
amplified during preheating \cite{BV} due to parametric resonance.

The power spectrum corresponding to (\ref{perturb}) is given by:
\begin{eqnarray}
{\cal P}_\Phi(k)=\frac{k^3 \vert C_k\vert ^2}{2\pi^2 V}=
A_S^2\left(\frac{k}{k_C}\right)^{n_s-1}
\label{power}
\end{eqnarray}
For simplicity we have assumed  a  power-law behaviour 
with spectral index $n_s$ and
we have set the normalization at the COBE scale 
$\lambda_C\simeq 3000 \;\mbox{Mpc}$ 
with $A_S\simeq 5 \cdot 10^{-5}$. Although there are some models
which predict this kind of behaviour, in general we will have a 
dependence of the spectral index on the scale. However, in
order to compare our results with those obtained from black
holes, we will keep this simple form.
In principle, there will be a maximum frequency cut-off $\kappa$
which corresponds to the  perturbation with the  smallest size 
produced during
inflation. Typically this scale is  roughly determined by 
the size of the horizon at the
end of inflation $\kappa\sim a_IH_I$
where the $I$ index denotes the end of
inflation  (for a detailed discussion see 
\cite{Bassett}).
We can obtain an explicit expression for the occupation number  
(\ref{total})  
in terms of the power spectrum. Taking the continuum limit 
$\sum_p\rightarrow (2\pi)^{-3/2}V\int d^3p$, we get:

\begin{eqnarray}
N_k=\sum_{\lambda\lambda'}\int \frac{dp\;d\Omega}{(2\pi)^{3/2}}p^2 
\cos^2(\theta/2)
\frac{\vert C_{k+p}\vert^2}{4k^2V} (p-k)^2\left\vert 
\int_{\eta_0}^{\eta_1} (i{\cal F}'+{\cal F}(p-k))d\eta\right\vert^2
\end{eqnarray}
where we have used $k\eta\ll 1$ which is valid for
super-Hubble modes and $\theta$ is the angle between  $\vec p$ and
the $z$ axis. 
The total comoving number density of gravitinos is given by
$N=\int d^3k N_k$, typically this integral is dominated by the upper limit
of integration, so that we can approximate
 $N\simeq \kappa^3 N_{\kappa}$. 
Notice again that we are assuming no enhancement of high-momentum 
modes during preheating as it happens for instance in two-field models
of inflation. It is then
enough to calculate $N_{\kappa}$:
\begin{eqnarray}
N_{\kappa}\simeq \frac{4\pi^3}{(2\pi)^{3/2}}\int dp\; p^2 
\frac{{\cal P}_\Phi(\kappa)}{\kappa^3}\simeq \frac{\sqrt{2\pi^3}}{3}
{\cal P}_\Phi(\kappa)
\label{nkgrav}
\end{eqnarray}
Because of Pauli exclusion principle $N_k\leq 1$, the violation
of this bound signals the breakdown of the perturbative approach.

In order to get the ratio $n/s$ we need to calculate the entropy density 
at the end of inflation. If we assume that all the energy density in the 
inflaton field is instantaneously converted into radiation, we get  
a very high reheating
temperature, which is already excluded by the
thermal production of gravitinos as commented before. Therefore, for  realistic
models, reheating and thermalization should occur sufficiently
late, so that we can have low reheating temperature. In that case, we must 
consider the stage of inflaton oscillations at the end of 
inflation. If the inflaton
potential close to the minimum behaves as $V\sim \phi^{\,\alpha}$, 
then the energy
density during oscillations scales as $\rho\sim a^{-6\alpha/(\alpha+2)}$ 
\cite{turner}.
Thus,  although the physical number density scales as $n\sim a^{-3}$,
the entropy density will do as $s\simeq \rho^{3/4}\sim 
a^{-9\alpha/(2(\alpha+2))}$.
This implies that during the oscillations and 
depending on the value of $\alpha$, the number density
$n$ can be diluted with respect to the entropy density, it could  
remain constant
or even increase. In the following, we will consider
the usual potentials with $\alpha=2,4$ (notice that we are approximating the
potential by simple powers only close to the minimum, during inflation
their behaviour can be completely different).    
For $\alpha=2$, the equation of state of the oscillating scalar
field can be considered as that of non-relativistic particles, 
in this case $n/s$ decreases in time. For $\alpha=4$, the oscillations 
behave as radiation and $n/s$ is constant. Thus, taking
this into account and setting $\kappa=a_IH_I$, 
 we get for the ratio at the end of reheating:
\begin{eqnarray}
\frac{n}{s} \simeq \frac {N_{\kappa} \;H_I^3}{T_R^3} 
\left(\frac{a_I}{a_R}\right)^3 \simeq  N_{\kappa} \left(\frac{H_I}{\sqrt{3}\;M_P}\right)^{3/2}
\left(\frac{T_R^2}{\sqrt{3}\;H_I\;M_P}\right)^{\frac{4-\alpha}{2\alpha}}
\label{ratio}
\end{eqnarray}
where $M_P=(8\pi G)^{-1/2}$. 
\begin{figure}[h]
\hspace{2cm}\mbox{\epsfysize=8cm\epsfxsize=12cm
\epsffile{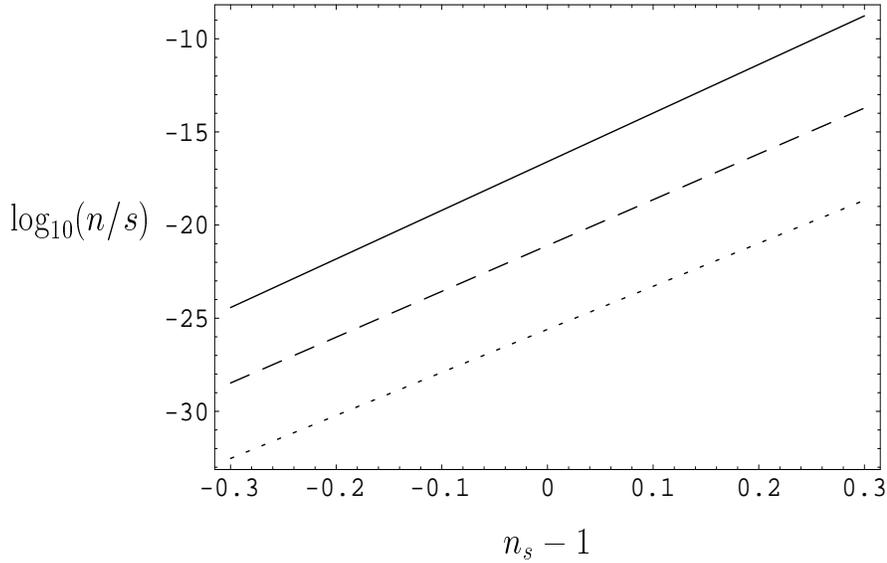}}
\caption{Abundance of gravitinos $n/s$ versus spectral index for a model
with $\alpha=4$ and 
$H_I=10^{13}$ GeV (full line), $H_I=10^{10}$ GeV (dashed line)
and $H_I=10^{7}$ GeV (dotted line). The results are independent of the 
reheating temperature}
\end{figure}
In order to compute $N_\kappa$ 
in the previous equation
 we need to know  the value of the ratio $\kappa/k_C$. 
If the Hubble parameter remains constant during inflation we have:
 $\kappa/k_C=a_I/a_C$, where $a_C$ denotes the scale factor at the moment
when the scale $\lambda_C$ left the horizon. Therefore, 
in terms of the number of e-folds of inflation, we can write:
\begin{eqnarray}
\frac{\kappa}{k_C}=e^{N(k_C)}
\end{eqnarray}
\begin{figure}[h]
\hspace{2cm}\mbox{\epsfysize=8cm\epsfxsize=12cm
\epsffile{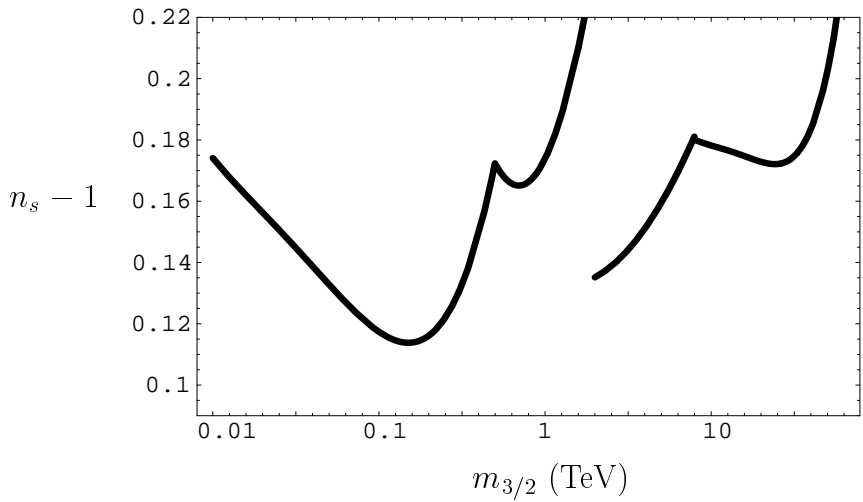}}
\caption{Limits on the spectral index $n_s$ as a function of the gravitino mass
$m_{3/2}$, from the generation of massive unstable gravitinos \cite{Ellis}.
The right curve corresponds to the hadroproduction of 
D and $^4$He \cite{Reno}. The left curve to the photofission of D and
the photoproduction of D and $^3$He 
\cite{Ellis2}. The model of inflation has $\alpha=4$ 
and $H_I=10^{13}$ GeV}
\end{figure}
\begin{figure}[h]
\hspace{2cm}\mbox{\epsfysize=8cm\epsfxsize=12cm
\epsffile{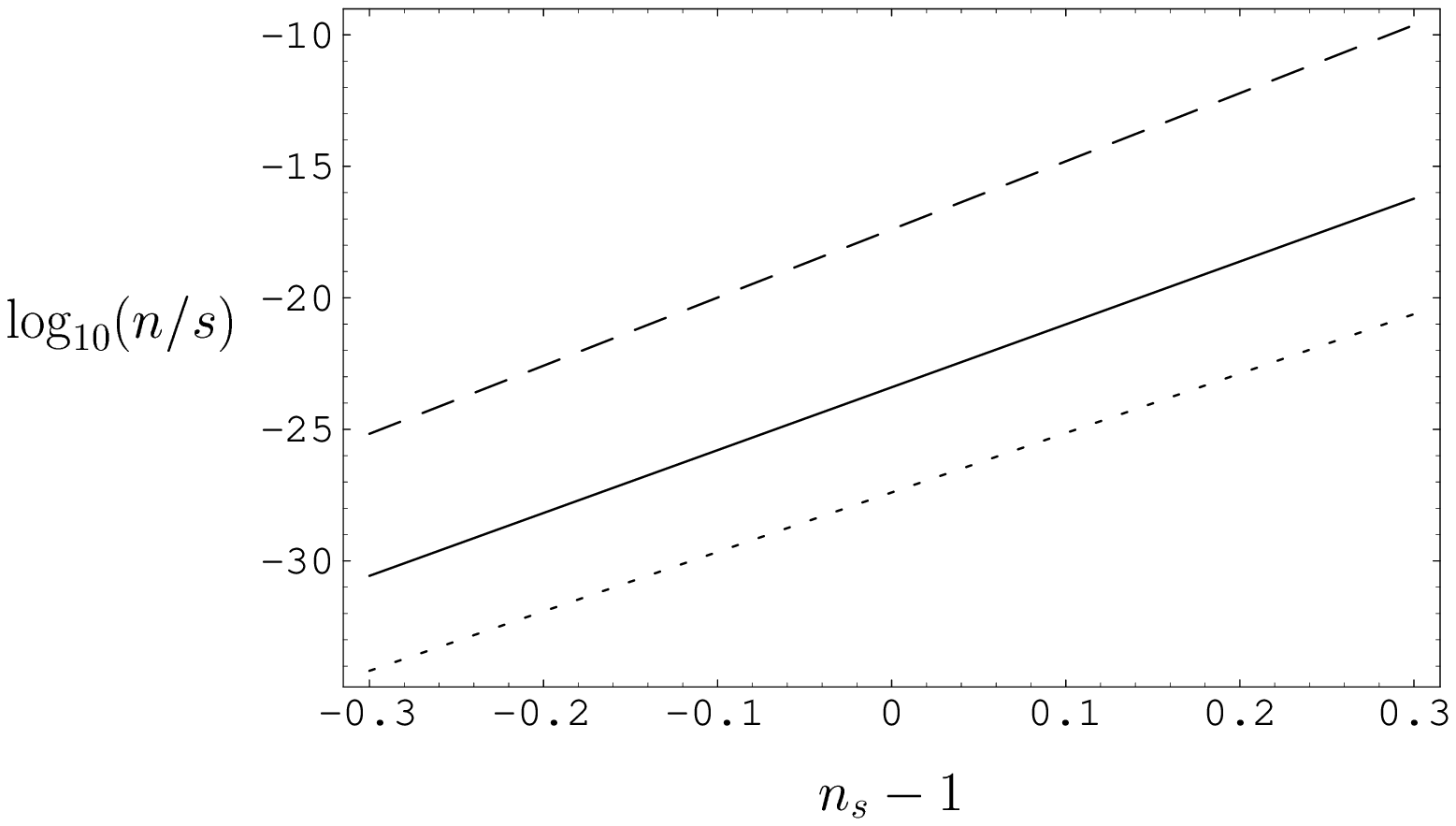}}
\caption{Abundance of gravitinos $n_{3/2}/s$ versus spectral index
$n_s$ for a model
with $\alpha=2$ and reheating temperature 
 $T_R=10^{9}$ GeV (full line), $T_R=10^{15}$ GeV (dashed line)
and $T_R=10^{5}$ GeV (dotted line), for $H_I=10^{13}$ GeV}
\end{figure}
In the case $\alpha=2$, $N(k_C)$  is given by the well-known expression 
\cite{Early}:
\begin{eqnarray}
N(k_C)=53+\frac{1}{3}
\ln\frac{T_R}{10^{10} GeV}+\frac{2}{3}\ln \frac{V^{1/4}}{10^{14} GeV} 
\end{eqnarray}
where $V$ is the value of the inflaton potential during inflation. 
A similar expression can be obtained for $\alpha=4$:
\begin{eqnarray}
N(k_C)=56+
\ln \frac{V^{1/4}}{10^{14} GeV} 
\end{eqnarray}
which is independent of the reheating temperature.
Substituting back in (\ref{ratio}) and using (\ref{power}) and (\ref{nkgrav}),
 we get:
\begin{eqnarray}
\frac{n}{s}
\simeq  \frac{\sqrt{2\pi^3} A_S^2}{3}
\left(\frac{H_I}{\sqrt{3}\;M_P}\right)^{3/2}
\left(\frac{T_R^2}{\sqrt{3}\;H_I\:M_P}\right)^{\frac{4-\alpha}{2\alpha}}
\exp({(n_s-1)N(k_C)})
\end{eqnarray}

In Fig.1 we have plotted $n/s$ as a function of the spectral 
index for different models with $H_I=10^{13},\; 10^{10}$ and $10^7$ GeV
with $\alpha=4$. In this case the results are independent of the reheating
temperature. The nucleosynthesis bounds on the spectral index
coming from the effects of the hadronic and radiative decays
of gravitinos on elemental abundances is summarized in Fig.2 for
a typical model with $H_I=10^{13}$ GeV. We see that the 
strongest bound comes from gravitino masses
around $100$ GeV for which $n_s<1.12$. This limit improves that 
imposed by primordial black holes production. In fact, for 
$T_R<10^9$ GeV, such limit comes from black holes
which are evaporating today, and is given by $n_s<1.28$ 
\cite{Liddle,Green}. 
Notice that for $n_s=1.12$ and 
$\kappa/k_C \sim 10^{26}$ 
(which is the range of scales spanned in the 
model considered) we have ${\cal P}_{\Phi}(\kappa)\simeq 3 \cdot
10^{-6}\ll 1$, i.e. we are well inside the perturbative
region. Notice that such perturbations 
are unable to create black holes, but still we  have an important effect
coming from the creation of relics.

In Fig.3 we plot the number density versus spectral index for a model 
with $\alpha=2$ and $H_I=10^{13}$ GeV, for different values of the reheating
temperature. Now because of the relative growth of entropy
during the inflaton oscillations, the production is much weaker than in the
previous case. In fact, in order to get a cosmologically relevant abundance
with low reheating temperature, the 
spectral index should be larger than $1.4$. 
However for those values  we find ${\cal P}_{\Phi}>1$, i.e. we are
out of the perturbative regime. For that reason for $\alpha=2$ 
we cannot obtain
limits on the spectral index. In Fig. 3 we have also plotted the
relative abundance for a model with high reheating temperature ($T_R=10^{15}$ 
GeV) with dashed line. The results show that even if we ignored the thermal
production of gravitinos, we will find again the problem, but now coming
from metric perurbations.

In \cite{Bassett} a  weak production  was obtained with numerical 
calculations for the $V=m^2\phi^2/2$ and $V=\lambda\phi^4/4$ models of
chaotic inflation. This is due to the fact that those models 
predict a power spectra
with negative tilt, i.e. $n_s<1$.

\section{Moduli production}

String theory and other higher-dimensional models include scalar
fields that parametrize the shape of extra dimensions and whose couplings
to matter fields are Planck mass suppressed. This kind of fields can give
rise to different cosmological problems \cite{Carlos}. First, if 
they were displaced from their minima in the early universe,  
we find the so called classical
moduli problem in which the energy density of the oscillations of the 
scalar fields behaves as non-relativistic matter, and can give relevant
contribution to the energy density of the universe. If these fields
decay during nucleosynthesis, we find a problem similar to that produced 
by gravitinos. The typical bounds on the primordial abundance of moduli
are also similar $n/s<10^{-12}-10^{-14}$ \cite{Ellis}. 
Second, moduli field can be
produced from quantum fluctuations during inflation. In this case,
the breaking of conformal invariance comes either from the minimal coupling
to the scalar curvature or from their own  mass. Thus the 
 typical equation of motion for moduli is given by:
\begin{eqnarray}
(\Box+m_\chi^2+\xi R)\chi=0
\end{eqnarray}
In general, the moduli mass acquires corrections, coming from supersymmetry
breaking effects during inflation or from non-renormalizable effects in
the superpotential, thus one has $m_{\chi}^2=m^2+C^2 H^2$, 
where $C$ is a model dependent parameter. 
\begin{figure}[h]
\hspace{2cm}\mbox{\epsfysize=8cm\epsfxsize=12cm
\epsffile{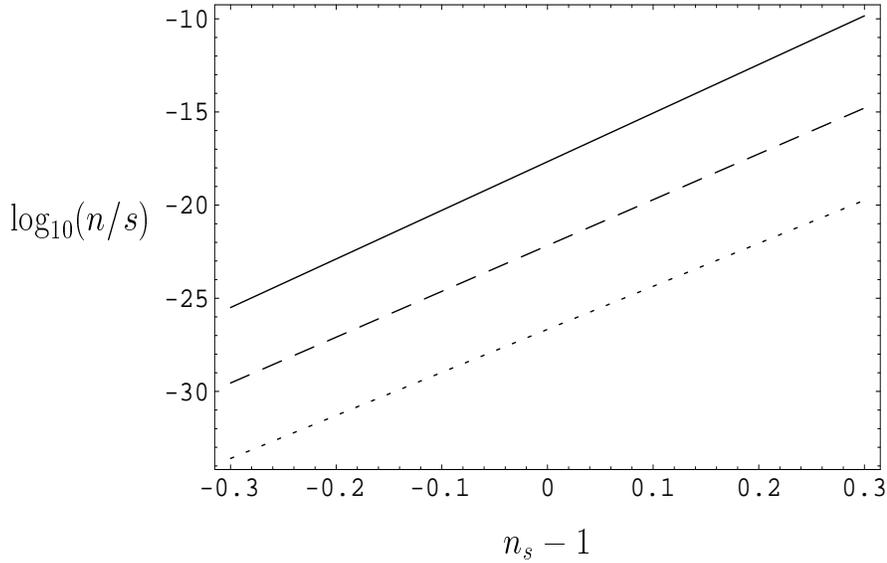}}
\caption{Abundance of moduli $n_{\chi}/s$ versus spectral index $n_s$ 
for a model
with $\alpha=4$ and $H_I=10^{13}$ GeV 
(full line), $H_I=10^{10}$ GeV (dashed line)
and $H_I=10^{7}$ GeV (dotted line). The results are independent of the 
reheating temperature.}
\end{figure}
\begin{figure}[h]
\hspace{2cm}\mbox{\epsfysize=8cm\epsfxsize=12cm
\epsffile{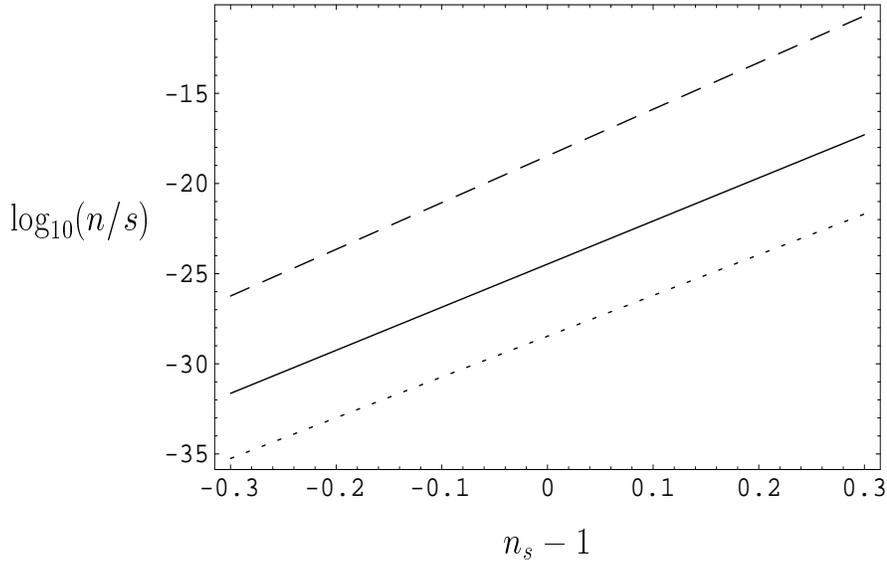}}
\caption{Abundance of moduli $n_{\chi}/s$ versus spectral index
$n_s$ for a model
with $\alpha=2$ and reheating temperature 
$T_R=10^{9}$ GeV (full line), $T_R=10^{15}$ GeV (dashed line)
and $T_R=10^{5}$ GeV (dotted line), for $H_I=10^{13}$ GeV.}
\end{figure}

It has been shown \cite{Lindemod} that the classical moduli problem is not
present in the cases $C\ll 1$ or $C\gg 1$. The gravitational 
quantum production of moduli could also be avoided if conformal
invariance is recovered, i.e., if $C=0$, $\xi=1/6$ and $m=0$,
although such fine tuning in the parameters seems to be difficult to obtain 
without invoking additional symmetries \cite{Riotto1}. Since we are interested
in the effect of metric perturbations on the moduli production, we will
consider in the following that they are the only source of conformal
symmetry breaking. In that case the above equation of motion reduces to:
\begin{eqnarray}
\left(\frac{1}{\sqrt{g}}\partial_\mu g^{\mu\nu}\sqrt{g}\partial_\nu
+\frac{1}{6}R\right)\chi=0
\end{eqnarray}
Using the form of the metric in (\ref{metric}) and  the formulae in the 
Appendix, we find  the following form of the linearized equation
of motion:
\begin{eqnarray}
\tilde \chi''-\partial_i\partial_i\tilde\chi-4\Phi'\tilde\chi'
-\frac{1}{3}(\partial_i\partial_i\Phi)\tilde\chi-\Phi''\tilde\chi=0
\end{eqnarray}
where $\tilde\chi=a\chi$. Again we look for solutions in the form
$\tilde \chi=\tilde \chi^{(0)}+\tilde \chi^{(1)}+\dots$ where
\begin{eqnarray}
\tilde \chi^{(0)}=\frac{1}{\sqrt{2pV}}e^{i(\vec p\vec x-p\eta)}
\end{eqnarray}
corresponding to the initial conformal vacuum state.
Fourier transforming the equation, we find for the first order correction:
\begin{eqnarray}
\tilde\chi^{(1)''}(\vec k, \eta)+k^2\tilde\chi^{(1)}(\vec k,\eta)+J=0
\end{eqnarray}
where
\begin{eqnarray}
J=\frac{1}{\sqrt{2pV}}\left(4ip \Phi'+\frac{1}{3}(\vec p+\vec k)^2\Phi
-\Phi''\right)e^{-ip\eta}
\end{eqnarray}

Using the analogous result to the gravitino case, we find for
the Bogolyubov coefficients:
\begin{eqnarray}
\beta_{pk}=-\frac{i}{\sqrt{2kV}}\int_{\eta_0}^{\eta_1} Je^{-ik\eta} d\eta
\end{eqnarray}
and finally for the occupation number we get:
\begin{eqnarray}
N_k =\int \frac{d^3p}{(2\pi)^{3/2}}\frac{\vert C_{p+k}\vert^2}{4kpV}\left\vert
\int_{\eta_0}^{\eta_1} \left(4ip{\cal F}'+\frac{1}{3}(\vec p+\vec k)^2{\cal F}-{\cal F}''\right)d\eta
\right\vert ^2
\end{eqnarray}

Assuming again that the total number density is dominated by those
modes with $k\sim \kappa$, we have:
\begin{eqnarray}
N_\kappa \simeq  \frac{32\pi^3}{(2\pi)^{3/2}\kappa}\int dp \frac{p^3}{\kappa^3}
{\cal P}_{\Phi}(\kappa)\left\vert \int_{\eta_0}^{\eta_1} \left(i{\cal F}'
+\frac{\kappa^2}{12p}{\cal F}
-\frac{{\cal F}''}{4p}\right)
d\eta\right\vert^2 \simeq \frac{(2\pi)^{3/2}}{72}{\cal P}_{\Phi}(\kappa)
\label{nkmod}
\end{eqnarray}
where the time integral has been estimated to be  
$\simeq\kappa/(12p)$, since ${\cal F}\simeq {\cal O}(1)$. 

The ratio $n/s$ can be calculated in a similar fashion using
(\ref{ratio}) and the expression we have just obtained for 
$N_\kappa$. The results are plotted in  Figs. 4,5 for different models 
with $\alpha=2,4$. In the case
in which there is no relative dilution of $n$ with respect to $s$, we obtain
that for a moderate value of the nucleosynthesis bound $n/s<10^{-13}$, 
the spectral index should be  $n_s<1.18$, whereas for the
strongest one $n/s<10^{-14}$, we get $n_s<1.14$ for 
$H_I\sim 10^{13}$ GeV. These limits are slightly
weaker than those coming from gravitinos, in part this is due to the 
absence of  spin degrees of freedom 
in this case. For $\alpha=2$ the limits
are much less stringent as expected. 

\section{Conclusions}
In this paper we have studied the production of helicity $\pm 3/2$ 
gravitinos and moduli fields from metric perturbations after inflation.
The results show that those metric perturbations with 
very  small wavelengths reentering the horizon right at the end of inflation
induce a strong deviation from conformal flatness which is responsible
for the production of a non-negligible amount of relics. In particular, 
if the power spectrum has a positive tilt $n_s>1$, as predicted by some
models of hybrid inflation \cite{hybrid}, the production could
conflict with the limits imposed by nucleosynthesis. This in turn allows
us to set stringent constraints on the spectral index in the particular
case in which $n_s$ does not depend on the scale. When 
the energy density in the inflaton oscillations scales as radiation, 
the limits
that we obtain $n_s<1.12$ are stronger than those coming from primordial
black hole production. In the case in which the energy in oscillations
scales as non-relativistic matter, the bounds are much weaker.
When the spectral index depends on the scale, 
it is possible to translate the nucleosynthesis bounds
into limits on the power spectrum at the smallest scale 
${\cal P}_{\Phi}(\kappa)$.
Using the results in the previous sections, we obtain
${\cal P}_{\Phi}(\kappa)<10^{-6}$, in
models of inflation with $H_I\sim 10^{13}$ GeV and $\alpha=4$.
In conclusion, we consider that the  
production of gravitinos and 
moduli fields provides us with a new tool which can be useful 
to study the region of very small wavelengths in the primordial
spectrum of metric peturbations.  

\vspace{1.7cm}

\centerline{\bf Acknowledgments}
I am grateful to A. Linde and L. Kofman for useful comments. 
This work has been partially supported by the CICYT (Spain) projects
AEN97-1693 and FPA2000-0956. The author  also acknowledges support from 
the Universidad Complutense del Amo Fellowship.

\vspace{.5cm}

\appendix
\section{Useful formulae}
In this Appendix we show some useful results for the perturbative expansions
of different geometrical objects, (see also \cite{Verdaguer}). 

\vspace{.5cm}

\centerline{\bf Vierbein}

The vierbein expansion corresponding to the metric tensor in (\ref{metric})
is given by:
\begin{eqnarray}
e^b_\mu&=&a\left(\eta^b_\mu+\frac{1}{2}h^b_{\;\mu}\right)
\end{eqnarray}
where
\begin{eqnarray}
h^0_{\;0}=2\Phi,\;\; 
h^i_{\;j}=-2\Phi\delta^{i}_{\;j}
\end{eqnarray}

\vspace{.5cm}

\centerline{\bf Gamma matrices}

The {\it curved} gamma matrices can be expanded as: 
\begin{eqnarray}
\gamma_\mu=e^a_\mu\hat\gamma_a=\gamma_\mu^{(0)}+\gamma_\mu^{(1)}+\cdots
\end{eqnarray}
where
\begin{eqnarray}
\gamma_0^{(1)}=a\Phi\hat\gamma_0=\Phi\gamma_0^{(0)},
\;\; \gamma_i^{(1)}=-a\Phi\hat\gamma_i=-\Phi\gamma_i^{(0)}
\end{eqnarray}

\centerline{\bf Spin connection}

For the spin connection appearing in the fermionic derivatives we find:
\begin{eqnarray}
\Gamma_\mu&=&\Gamma_\mu^{(0)}+\Gamma_\mu^{(1)}+\dots\nonumber\\
\Gamma_\mu^{(0)}&=&\frac{a_{,\lambda}}{4a}\eta^{b\lambda}\eta^a_{\;\mu}
[\hat\gamma_a,\hat\gamma_b]\nonumber\\
\Gamma_0^{(0)}&=&0,\;\;
\Gamma_j^{(0)}=\frac{a'}{2a^3}\gamma_j^{(0)}\gamma_0^{(0)}
\nonumber \\
\Gamma_\mu^{(1)}&=&\frac{1}{8}[\hat\gamma_a,\hat\gamma_b]
\left(h_\mu^{\;a,b}+\eta^b_{\;\mu}h^{a\lambda}\frac{a_{,\lambda}}{a}
+\eta^{b\lambda}h_\mu^{\;a}\frac{a_{,\lambda}}{a}\right)\nonumber\\
\Gamma_0^{(1)}&=&\frac{1}{2}\gamma_0^{(0)}\gamma^{(0)j}\Phi_{,j} \nonumber\\
\Gamma_j^{(1)}&=&-\frac{1}{4}\Phi_{,\mu}[\gamma_j^{(0)},\gamma^{(0)\mu}]
+\Phi\frac{a'}{a^3}\gamma_0^{(0)}\gamma_j^{(0)}
\end{eqnarray}

\centerline{\bf Christoffel symbols}

The non-vanishing Christoffel symbols for the metric (\ref{metric}) 
are given by:
\begin{eqnarray}
\Gamma^{\lambda}_{\mu\nu}&=&\Gamma^{(0)\lambda}_{\;\;\;\;\;\mu\nu}+
\Gamma^{(1)\lambda}_{\;\;\;\;\;\mu\nu}
+\dots\nonumber\\
\Gamma^{(0)0}_{\;\;\;\;\;00}&=&\frac{a'}{a},\;\;\Gamma^{(0)0}_{\;\;\;\;\;ij}=
\frac{a'}{a}\delta_{ij},\nonumber \\ 
\Gamma^{(0)i}_{\;\;\;\;\;0j}&=&\frac{a'}{a}\delta^i_{\;j}\nonumber \\
\Gamma^{(1)0}_{\;\;\;\;\;00}&=&\Phi',\;\; 
\Gamma^{(1)0}_{\;\;\;0i}=\Phi_{,i}\nonumber\\
\Gamma^{(1)0}_{\;\;\;\;\;ij}&=&
\left(-4\frac{a'}{a}\Phi-\Phi'\right)\delta_{ij},\;\;
\Gamma^{(1)i}_{\;\;\;\;\;0j}=-\Phi'\delta^i_{j}\nonumber \\
\Gamma^{(1)i}_{\;\;\;\;\;jk}&=
&-\left(\Phi_{,k}\delta^i_{\;j}+\Phi_{,j}\delta^i_{\;k}
-\Phi_{,l}\delta_{jk}\delta^{il}\right)\nonumber \\
\Gamma^{(1)i}_{\;\;\;\;\;00}&=&\Phi_{,i}
\end{eqnarray}

\centerline{\bf Scalar curvature}

Finally, for the scalar curvature we find up to first order:
\begin{eqnarray}
R=\frac{1}{a^3}\left[6a''-12\Phi\;a''-24a'\;\Phi'
+2a\partial_i\partial_i\Phi-6\Phi''\;a\right]\nonumber\\
\end{eqnarray}

\centerline{\bf Spinors}

The normalized spinors with helicity $\pm 1/2$ and
momentum $\vec p$ that we have used in the text are:
\begin{eqnarray}
u(\vec p,\pm)&=&\sqrt{p}
\left(
\begin{array}{c}
\chi_\pm\\ \pm \chi_\pm
\end{array}
\right),\;\;
 \\
\chi_+=\left(
\begin{array}{c}
e^{-i\phi/2}\cos\left(\frac{\theta}{2}\right)\\
e^{i\phi/2}\sin\left(\frac{\theta}{2}\right) 
\end{array}
\right)&,&\;\;
\chi_-
=\left(\begin{array}{c}
-e^{-i\phi/2}\sin\left(\frac{\theta}{2}\right)\\
e^{i\phi/2}\cos\left(\frac{\theta}{2}\right) \nonumber
\end{array}
\right)
\end{eqnarray}


\end{document}